\documentclass[twocolumn,showpacs,preprintnumbers,amsmath,amssymb]{revtex4}
\usepackage{graphicx}
\usepackage{dcolumn}
\usepackage{bm}
\begin{document}
\newcommand{\hs}{\hspace*{0.5cm}}
\newcommand{\vs}{\vspace*{0.5cm}}
\newcommand{\be}{\begin{equation}}
\newcommand{\ee}{\end{equation}}
\newcommand{\bea}{\begin{eqnarray}}
\newcommand{\eea}{\end{eqnarray}}
\newcommand{\ben}{\begin{enumerate}}
\newcommand{\een}{\end{enumerate}}
\newcommand{\bde}{\begin{widetext}}
\newcommand{\ede}{\end{widetext}}
\newcommand{\nn}{\nonumber}
\newcommand{\crn}{\nonumber \\}
\newcommand{\non}{\nonumber}
\newcommand{\noi}{\noindent}
\newcommand{\al}{\alpha}
\newcommand{\la}{\lambda}
\newcommand{\bet}{\beta}
\newcommand{\ga}{\gamma}
\newcommand{\va}{\varphi}
\newcommand{\om}{\omega}
\newcommand{\pa}{\partial}
\newcommand{\fr}{\frac}
\newcommand{\bc}{\begin{center}}
\newcommand{\ec}{\end{center}}
\newcommand{\Ga}{\Gamma}
\newcommand{\de}{\delta}
\newcommand{\De}{\Delta}
\newcommand{\ep}{\epsilon}
\newcommand{\varep}{\varepsilon}
\newcommand{\ka}{\kappa}
\newcommand{\La}{\Lambda}
\newcommand{\si}{\sigma}
\newcommand{\Si}{\Sigma}
\newcommand{\ta}{\tau}
\newcommand{\up}{\upsilon}
\newcommand{\Up}{\Upsilon}
\newcommand{\ze}{\zeta}
\newcommand{\ps}{\psi}
\newcommand{\Ps}{\Psi}
\newcommand{\ph}{\phi}
\newcommand{\vph}{\varphi}
\newcommand{\Ph}{\Phi}
\newcommand{\Om}{\Omega}
\def\lappeq{\mathrel{\rlap{\raise.5ex\hbox{$<$}}
{\lower.5ex\hbox{$\sim$}}}}

\preprint{KEK-TH-1223}

\title{Neutrino masses and lepton flavor violation in the 3-3-1 model
with right-handed neutrinos}

\author{P. V. Dong} \email{pvdong@iop.vast.ac.vn} \altaffiliation[On leave of absent from ]{Institute
of Physics, VAST, Vietnam} \affiliation{Theory Group, KEK, 1-1 Oho,
Tsukuba, 305-0801, Japan}

\author{H. N. Long} \email{hnlong@iop.vast.ac.vn}
\affiliation{Institute of Physics,
VAST, P.O. Box 429, Bo Ho, Hanoi 10000, Vietnam}

\date{\today}

\begin{abstract}
We show that in the framework of the 3-3-1 model with right-handed
neutrinos, small neutrino masses and large lepton flavor violating
processes such as $\mu\rightarrow 3 e$ and $\mu\rightarrow e\gamma$
can be obtained by just introducing an additional Higgs sextet. In
the limit of vanishing of the Yukawa interaction among Higgs and
lepton triplets $(h^\nu=0)$, the decay $\mu\rightarrow 3 e$ strongly
depends on the neutrino mass patterns, but the $\mu\rightarrow
e\gamma$ almost does not. The neutrino masses are not constrained by
such processes in the cases of $h^\nu\neq 0$.
\end{abstract}

\pacs{14.60.Pq, 13.35.Bv, 12.60.Fr, 12.60.Cn}

\maketitle

\section{\label{intro}Introduction}

The recent experimental results confirm that neutrinos have tiny
masses and oscillate \cite{pdg}, this implies that the standard
model (SM) must be extended. Among the beyond-SM extensions, the
models based on the $\mathrm{SU}(3)_C\otimes \mathrm{SU}(3)_L
\otimes \mathrm{U}(1)_X$ (3-3-1) gauge group \cite{ppf,flt} have
some intriguing features: First, they can give partial explanation
of the generation number problem. Second,  the third quark
generation has to be different from the first two, so this leads to
the possible explanation of why top quark is uncharacteristically
heavy. An additional motivation to study this kind of the models is
that they can also predict the electric charge quantization
\cite{prs}.

Such 3-3-1 models have been studied extensively over the last
decade. In one of them the three lepton triplets are of the form
$(\nu, l, \nu^c)_L$, where the right-handed (RH) neutrinos
$\nu_R^c=(\nu^c)_L$ are included into the third components of
triplets. Traditionally, this model works with the three Higgs
triplets and named the 3-3-1 model with RH neutrinos \cite{flt}. At
the tree level the neutrino spectrum contains three Dirac fields
with one massless and two degenerate in mass~$\sim h^\nu v$, where
the $v$ vacuum expectation value (VEV) is related to the electroweak
scale, and the Majorana fields $\nu_L$ and $\nu_R$ are massless.
This spectrum is not realistic under the data because there is only
one squared-mass splitting. Since the observed neutrino masses are
so small, the Dirac mass is unnatural, and one must understand what
physics gives $h^\nu v \ll h^l v$---the mass of charged leptons. The
neutrino masses are thus a great question addressed to this kind of
the models.

Alternatively, the neutrino oscillation shows clearly that the
lepton flavors are not conserved in nature. If we accommodate this
feature simply by an introduction of neutrino masses in the SM,
other lepton flavor violating (LFV) processes such as
$\mu\rightarrow e\gamma$ would still have so small a rate (branching
ratio $\leq10^{-40}$) that there is no hope for their detection in
the foreseeable future \cite{chli,chli1}. This is also the case of
the model under consideration. The LFV processes such as
$\mu\rightarrow e\gamma$ and $\mu\rightarrow 3e$ found are very
suppressed too. In this work, we are particularly interested in the
possibility of generating small neutrino masses so that the LFV
processes are less damped. We stress that this might be achieved by
introducing just an additional Higgs sextet. This sextet now becomes
a nice element because the neutrino mass is the result of a type II
seesaw \cite{typeIIseesaw} and the LFV processes in a case are
mediated only by the doubly-charged scalar \cite{kuno-okada}. The
sterile neutrinos mix small with the ordinary ones and have masses
in the 3-3-1 symmetry breaking scale.

This work is organized as follows. In Sec. \ref{model} we present
the model. Section \ref{higgs-neutrino} introduces the Higgs sextet
and give a realistic mass spectrum for the neutrinos. Section
\ref{lfv} is devoted to the LFV processes $\mu\rightarrow 3e$ and
$\mu\rightarrow e\gamma$. Phenomenological relationship among the
neutrino spectra and LFV decays is obtained in the end of this
section. We make conclusions in Section \ref{conclus}.

\section{\label{model} A review of the model}

The model under consideration is of the 3-3-1 model with RH
neutrinos \cite{flt}. The particle content which is anomaly free is
given as follows $\psi_{aL}=(\nu_{aL},l_{aL},\nu^c_{aR})^T \sim (3,
-1/3)$ $(a = 1, 2, 3)$, $l_{aR}\sim (1, -1)$,
$Q_{3L}=(u_{3L},d_{3L}, U_{L})^T\sim (3,1/3)$, $Q_{\al L}=(d_{\al
L},  -u_{\al L},  D_{\al L})^T\sim (3^*,0)$ $(\al=1,2)$, $u_{aR}\sim
(1,2/3)$, $d_{a R} \sim (1,-1/3)$, $U_{R}\sim (1,2/3)$, $D_{\al R}
\sim (1,-1/3)$. The values in the parentheses denote quantum numbers
based on the $\left(\mbox{SU}(3)_L,\mbox{U}(1)_X\right)$ symmetry.
The electric charge operator takes a form
$Q=T_3-\fr{1}{\sqrt{3}}T_8+X$, where $T_i$ $(i=1,2,...,8)$ and $X$
stand for $\mbox{SU}(3)_L$ and $\mbox{U}(1)_X$ charges,
respectively. The electric charges of exotic quarks are the same as
of the ordinary ones: $q_{U}=\fr 2 3$ for $U$ and $q_D=-\fr 1 3$ for
$D_\al$.

The electroweak symmetry breaking in this model is through two
stages, $\mathrm{SU}(3)_L\otimes \mathrm{U}(1)_X \rightarrow
\mathrm{SU}(2)_L\otimes\mathrm{U}(1)_Y \rightarrow \mathrm{U}(1)_Q$,
achieved by three Higgs triplets: $\chi = (\chi^0_1, \chi^-_2,
\chi^{0}_3 )^T \sim (3, -1/3)$, $\eta=(\eta^0_1, \eta^-_2,
\eta^{0}_3 )^T\sim (3, -1/3)$, $\rho =(\rho^+_1, \rho^0_2,
\rho^{+}_3 )^T \sim (3, 2/3)$, with the VEVs corresponding to
$\langle\chi \rangle = (u'/\sqrt{2}, 0,w/\sqrt{2})^T$, $\langle\eta
\rangle  = ( u/ \sqrt{2}, 0, w'/\sqrt{2})^T$, $\langle\rho \rangle =
(0, v/ \sqrt{2}, 0)^T$. The most general Yukawa Lagrangian
responsible for fermion masses is separated into two parts: \bea
{\mathcal L}_{\mathrm{LNC}}&=&h^U\bar{Q}_{3L}\chi
U_{R}+h^D_{\al\beta}\bar{Q}_{\al L}\chi^* D_{\beta R} +h^u_a
\bar{Q}_{3L}\eta u_{aR}\crn &&+h^d_{\al a}\bar{Q}_{\al L}\eta^*
d_{aR} +h^d_{a}\bar{Q}_{3 L}\rho d_{a R}+h^u_{\al a}\bar{Q}_{\al
L}\rho^* u_{aR}\crn && +h^l_{ab}\bar{\psi}_{aL}\rho
l_{bR}+h^\nu_{ab}\bar{\psi}^c_{aL}\psi_{bL}\rho+
\mathrm{H.c.},\label{y1}\\ {\mathcal
L}_{\mathrm{LNV}}&=&s^u_{a}\bar{Q}_{3L}\chi u_{aR}+s^d_{\al
a}\bar{Q}_{\al L}\chi^* d_{a R}+s^U\bar{Q}_{3L}\eta U_R\crn
&&+s^D_{\al \bet}\bar{Q}_{\al L}\eta^* D_{\bet R} +s^D_{
\al}\bar{Q}_{3L}\rho D_{\al R}+s^U_{\al }\bar{Q}_{\al L}\rho^*
U_{R}\crn &&+ \mathrm{H.c.},\label{y2}\eea where the subscripts LNC
and LNV respectively indicate to the lepton number conserving and
violating ones as shown below.

The Yukawa couplings of (\ref{y1}) possess an extra global symmetry
which is not broken by $u, v, w$ but by $u',w'$. From these
couplings, one can find the following lepton symmetry: $L(\nu_{aL},
l_{aL,R}, \nu_{aR})=1$, $L(\chi^{0*}_1, \chi^+_2, \rho^+_3,
\eta^0_3, U_{L,R}, D^*_{\alpha L,R})=-2$, and $L=0$ for other
fields. Here the $L$ is broken by $u'$ and $w'$ due to
$L(\chi^0_1,\eta^0_3)\neq 0 $. It is interesting that the exotic
quarks also carry the lepton number, so called leptoquarks. This $L$
obviously does not commute with the gauge symmetry, one can
construct a new conserved charge $\cal L$ through $L$ by making a
linear combination $L= xT_3 + yT_8 + {\cal L} I$. Applying $L$ on a
lepton triplet, the coefficients will be determined $L =
\fr{4}{\sqrt{3}}T_8 + {\cal L} I$,  where $\mathcal{L}(\chi)=4/3$,
$\mathcal{L}(\eta, \rho, Q_{3L}, Q^*_{\al L})=-2/3$,
$\mathcal{L}(u_{aR}, d_{aR})=0$, $\mathcal{L}(U_R, D^*_{\al R})=-2$,
$\mathcal{L}(\psi_{aL})=1/3$, and $\mathcal{L}(l_{aR})=1$. Another
useful conserved charge $\cal B$ exactly not broken by any VEV is
usual baryon number: $B ={\cal B} I$, where $\mathcal{B}=1/3$ for
all the quark multiplets and $\mathcal{B}=0$ for Higgs and lepton
ones. It is noteworthy that although $\chi$ and $\eta$ have the same
quantum numbers, they are discriminative due to difference in
$\mathcal{L}$-charge.

The interactions (\ref{y2}) violate $\mathcal{L}$ with $\pm 2$
units. In addition they imply mixing among the exotic quarks and
ordinary quarks of the same charge: $(u_1,u_2,u_3,U)$ and
$(d_1,d_2,d_3,D_1,D_2)$, this would lead to the flavor-changing
neutral-current processes \cite{mpp}. By those reasons, it should be
noted that the Yukawa couplings (\ref{y2}) must be respectively much
smaller than the first ones (\ref{y1}), $s\ll h$. Also, the lepton
number breaking VEVs are respectively much smaller than the usual
ones, $u'\ll u$ and $w'\ll w$. In this case all the quarks get mass
at the tree level. The Lagrangian (\ref{y2}) has often been excluded
commonly by the adoption of an appropriate discrete symmetry
\cite{mpp,ponce}, but there is no reason within the 3-3-1 models why
such Lagrangian should not be present.

The mass matrix for the charged leptons is obtained by
$M_l=-\fr{1}{\sqrt{2}}vh^l$, which is the same as in the ordinary
version \cite{flt}. Hereafter, we will assume that $h^l$ is flavor
diagonal, thus $l_{a}$ are mass eigenstates with respective to mass
eigenvalues $m_{a}=-\fr{1}{\sqrt{2}}v h^l_{aa}$.

The Lagrangian for the tree-level neutrino masses is obtained as\bea
\mathcal{L}^\mathrm{\nu}_\mathrm{mass}=-(M_D)_{ab}\bar{\nu}_{aR}\nu_{bL}
+\mathrm{H.c.},\hs M_D=-\sqrt{2}v h^\nu, \label{eq1}\eea where
$h^\nu_{ab}$ is antisymmetry in $a$ and $b$ due to Fermi statistics.
The tree-level spectrum therefore consists of only Dirac neutrinos
with one particle massless and two others degenerate in mass:
$0,-m,m$. This spectrum is unrealistic, but it could be severely
changed by the quantum corrections. In this case both the Dirac and
Majorana mass types get the possible corrections.

If such a tree-level spectrum dominates resulting masses after the
corrections, the model provides a possible explanation of the large
splitting $\Delta m^2_\mathrm{atm} \gg \Delta m^2_\mathrm{sol}$. But
we then must need a fine-tuning $m \sim (\Delta
m^2_\mathrm{atm})^{1/2}\sim 5\times10^{-2}\ \mathrm{eV}$, thus
$h^\nu\sim 10^{-13}$ \cite{pdg}. The coupling $h^\nu$ is so small
and therefore this fine-tuning is unnatural. Exactly, $h^\nu$ enter
the radiative corrections for the neutrino Majorana masses, these
masses would get so small values not large enough to split the
degenerate masses into a realistic spectrum (the largest splitting
in squared-mass is still much smaller than $\Delta
m^2_\mathrm{sol}\sim 8\times10^{-5}\ \mathrm{eV}^2$ \cite{pdg}). In
this case, the Dirac masses get trivially corrections \cite{dls1}.
Those conclusions are in {\it contradiction} with the previous one
given in \cite{changlong}.

We could introduce an appropriate discrete symmetry to suppress the
tree-level neutrino masses so that the neutrino spectrum is entirely
induced by either radiative corrections \cite{ponce} or effective
operators \cite{diasalex}. In such cases, if one takes an
examination on the LFV processes such as $\mu\rightarrow e\gamma$
and $\mu\rightarrow 3e$, the branching ratios are found to be very
suppressed (cf. \cite{chli}). If any detection of such processes is
positive, the models are thus no longer favored. In the following we
will search for the possibility of large LFV processes while still
keeps naturally-small neutrino masses.

\section{\label{higgs-neutrino} Higgs sextet and neutrino mass}

The neutrino Dirac masses by their naturalness will be treated as
large as of the usual charged ones resulting from the standard
symmetry breaking. We shall solve the problems as mentioned by
introducing just an additional Higgs sextet into the model. The
sextet has been considered formerly in \cite{vasi,mpp} in order to
accommodate for the neutrino masses, but the correct formulation of
neutrino-mass matrix has {\it not} been given. The physics
associated with this sextet has not yet been explored. New
interesting phenomenologies concerning the neutrino mass spectra and
LFV processes are to be studied in this work.

Decomposing $\bar{\psi}^c_{aL}\psi_{bL}\sim (3^*+6,-2/3)$, the
values $(3^*,-2/3)$ are just charges coupled to $\rho^*$. The
remaining $(6,-2/3)$ has been leaved before and now assigned to a
Higgs sextet. This sextet and its VEVs are, respectively, defined by
\be S=\left(
\begin{array}{ccc}
S^0_{11} & S^-_{12} & S^0_{13} \\
S^-_{12} & S^{--}_{22} & S^-_{23} \\
S^0_{13} & S^-_{23} & S^0_{33} \\
\end{array}
\right),\hs \langle S \rangle = \fr{1}{\sqrt{2}}\left(
\begin{array}{ccc}
\kappa & 0 & \vartheta \\
0 & 0 & 0 \\
\vartheta & 0 & \La \\
\end{array}
\right).
 \ee It has the following Yukawa
couplings: \be
\mathcal{L}_S=f^\nu_{ab}(\bar{\psi}^c_{aL})_m(\psi_{bL})_n(S^*)_{mn}
+\mathrm{H.c.}, \label{eq2} \ee where $f^\nu_{ab}$ is symmetry in
$a$ and $b$. The $S^{--}_{22}$ is the unique particle in the model
carrying the exotic charge ``$--$'', it does not mix with other
particles and hence becomes a physical scalar with mass $M$. One can
check the lepton charge $\mathcal{L}(S)=\fr 2 3$, thus
$L(S^0_{33})=-2$ and $L(S^0_{11},S^-_{12},S^{--}_{22})=2$ (other
components have vanishing $L$). The charged $S^-_{12},\ S^{--}_{22}$
are bilepton particles, and the neutral components $S^0_{11},\
S^0_{33}$ with respective to VEVs $\kappa$, $\La$ beak the lepton
number responsible for the left- and right-handed neutrino Majorana
masses, respectively.

Decomposing the Higgs multiplets into the SM ones, we get four
doublets $(\chi^0_1,\chi^-_2)^T$, $(\eta^0_1,\eta^-_2)^T$,
$(\rho^+_1,\rho^0_2)^T$ and $(S^0_{13},S^-_{23})^T$, a triplet
$(S^0_{11},S^-_{12},S^{--}_{22})$, and four singlets $\chi^0_3$,
$\eta^0_3$, $\rho^+_3$ and $S^0_{33}$. The VEVs of the singlet
components $w$ and $\La$ give mass for exotic quarks and RH Majorana
neutrinos as well as the new gauge bosons. The VEVs of doublets and
triplet give mass for all the ordinary fermions and gauge bosons. To
keep a consistency with the effective theory, including the
conditions as given previously, it is safe to impose the
constraints: $w'\ll w$ and $u',\kappa \ll u,v,\vartheta \ll w,\La$.
In this effective limit, the mass of $W$ boson and the
$\rho$-parameter are evaluated by $M^2_W \simeq
\fr{g^2}{4}(u^2+v^2+2\vartheta^2)$, $\rho=\fr{M^2_W}{c^2_W
M^2_Z}\simeq 1-\fr{2\kappa^2}{u^2+v^2+2\vartheta^2}$. We therefore
identify $u^2+v^2+2\vartheta^2=v^2_\mathrm{weak}\simeq (246\
\mathrm{GeV})^2$, and then obtain the limit $|\kappa|<2.46\
\mathrm{GeV}$ with the help of the data $\rho>0.9998$ \cite{pdg}.

With the aid of (\ref{y1}) and (\ref{eq2}), the Lagrangian
(\ref{eq1}) is rewritten in the form: \be
\mathcal{L}^\mathrm{\nu}_\mathrm{mass}=-\fr 1 2 \left(\bar{\nu}^c_L,
\bar{\nu}_R\right)M_\nu \left(
\begin{array}{c}
\nu_L \\
\nu^c_R \\
\end{array}
\right)+\mathrm{H.c.}, \ee where the mass matrix for the neutrinos
is obtained as follows \bea M_\nu=-\sqrt{2}\left(
   \begin{array}{cc}
     \kappa f^\nu & (vh^\nu+\vartheta f^\nu)^T \\
     vh^\nu+\vartheta f^\nu & \La f^\nu \\
   \end{array}
\right).\label{mm}\eea Let us remind the reader that the first term
of the Dirac mass matrix, $vh^\nu$, was {\it excluded} in the latter
one of \cite{vasi}. Because $\kappa\ll v,\vartheta\ll\La$, the
active neutrinos $\sim\nu_L$ gain mass via a type II seesaw: \bea
M_1 &\simeq& -\sqrt{2}\left\{\left(\kappa
-\fr{\vartheta^2}{\La}\right)
f^\nu-\fr{v^2}{\La}h^\nu(f^\nu)^{-1}h^\nu\right\}.\label{seesaw}\eea
It turns out that the neutrino masses in this model are naturally
small because of suppression of a large $\La$ scale and a small
$\kappa$ constrained from the $\rho$-parameter. The sterile
neutrinos $\sim \nu_R$ have large masses in the $\La$ scale:
$M_2\simeq -\sqrt{2}\La f^\nu$. Because $f^\nu$ and $h^\nu$ are,
respectively, symmetry and antisymmetry, the Dirac $vh^\nu+\vartheta
f^\nu$ in (\ref{mm}) is an arbitrary complex matrix. This means that
the seesaw mechanism as given in the model is quite general. As
shown below the coupling $f^\nu$ is more constrained by the LFV
processes but $h^\nu$ does not. This actually allows us to recover
appreciate neutrino-mass spectra by $h^\nu$ while keep the LFV
branching ratios in the present bounds.

Let us give a numerical estimation. Putting $u\sim v \sim \vartheta
\sim 100\ \mathrm{GeV}$ from the $W$ mass, $\kappa \Lambda \sim
\vartheta^2$, $f^\nu \sim h^\nu$, and $M_1\sim 1\ \mathrm{eV}$, the
sterile masses are proportional to $M_2\sim (h^\nu)^2
\fr{v^2}{M_1}\sim (h^\nu)^2 \times 10^{13}\ \mathrm{GeV}$. We see
that the seesaw scale $M_2\sim 1\ \mathrm{TeV}$ if $h^\nu$ is in
order of the electron Yukawa coupling $h^\nu\sim 10^{-5}$, and
$M_2\sim 10^8\ \mathrm{GeV}$ if $h^\nu$ rises to the muon or tauon
Yukawa coupling. For simplicity, in the following we will take the
last or higher value of $M_2$ into account, i.e. the contribution to
$\mu\rightarrow e\gamma$ shall entirely come from the doubly-charged
scalar and the values $f^\nu\sim 10^{-3}-10^{-2}$ follow.

\section{\label{lfv} Lepton flavor violating processes}

In this model, the process $\mu\rightarrow 3e$ is given at a tree
level diagram as mediated by the doubly-changed scalar. The
branching ratio is obtained by \be \mathrm{Br}(\mu\rightarrow
3e)\simeq \fr{\Ga(\mu\rightarrow 3e)}{\Ga(\mu\rightarrow e
\tilde{\nu}_e\nu_\mu)}\simeq \fr{1}{4
G^2_F}\fr{|f^{\nu\dagger}_{11}f^\nu_{12}|^2}{ M^4},\label{brm3e}\ee
where $G_F=1.16637\times 10^{-5}\ \mathrm{GeV}^{-2}$ is the Fermi
constant \cite{pdg}. Taking $M\sim (200-1000)\ \mathrm{GeV}$ and
$|f^{\nu\dagger}_{11} f^\nu_{12}|=10^{-6}$ as mentioned above, we
get $\mathrm{Br}(\mu\rightarrow 3e)\sim 10^{-12}-1.8\times 10^{-15}$
which {\it coincides} with the current experimental bound:
$\mathrm{Br}(\mu\rightarrow 3e) \leq 10^{-12}$ \cite{pdg}.

The process $\mu\rightarrow e\gamma$ is given at one-loop level
diagrams as mediated by the charged leptons $(e,\mu,\tau)$ and
doubly-charged scalar. There are two kinds of the diagrams among
them corresponding to photon emission from internal scalar and
fermion lines which yield relevant amplitudes. Other diagrams with
photon emission from external charged lines contribute only to the
$\bar{e}\ga^\la \mu\ep^*_\la$ amplitude which vanishes because of
current conservation. The branching ratio is given by \be
\mathrm{Br}(\mu\rightarrow e\ga)\simeq\fr{\al}{3\pi
G^2_F}\fr{|(f^{\nu\dagger}f^\nu)_{12}|^2}{M^4},\label{amp}\ee where
$\al=e^2/4\pi=1/128$ \cite{pdg}. Taking
$|(f^{\nu\dagger}f^\nu)_{12}|=10^{-4}$ and $M=(200-1000)$ GeV as
mentioned, we obtain $\mathrm{Br}(\mu\rightarrow e\ga)\sim 3.8\times
10^{-11}-6\times 10^{-14}$ which is {\it comparable} to the bound of
current experiments: $\mathrm{Br}(\mu\rightarrow e\ga) \leq 1.2
\times 10^{-11}$ \cite{pdg}.

At this one-loop level, there are other sources contributing to
$\mu\rightarrow e\gamma$. The first one is mediated by charged gauge
bosons---the SM $W^{\pm}$ and bilepton $Y^{\pm}$ because the
neutrinos mix, but these contributions are so small and thus safely
neglected \cite{chli}. The second comes from physical singly-changed
scalars similarly to the case of the doubly-charged scalar, but now
the internal fermion lines are neutrinos (in more details, see
\cite{chli1}). The contribution is quite smaller than those given in
(\ref{amp}) and can neglect if $M_1$ is in eV, $M_2$ of order
$\mathcal{O}(10^8-10^{16})$ GeV, and the singly-charge scalar masses
are the same order of doubly-charged one: $\mathcal{O}(100-1000)$
GeV. In another scenario the seesaw scale $M_2$ is in TeV, that
contribution is comparable to (\ref{amp}). Without loss of
generality we do not consider the case in this work because we are
interested only in the LFV size of the decay.

The relationship among neutrino mass spectra and LFV rates can be
divided into: {\it Case 1}. $h^\nu=0$: The mass matrix of neutrinos
(\ref{seesaw}) is rewritten as $M_1 =-\sqrt{2}\left(\kappa
-\fr{\vartheta^2}{\La}\right) f^\nu$. Thus the neutrino masses and
LFV muon decays depend only on $f^\nu$ which is similar to Higgs
triplet model \cite{mrs,mrs1}, but in our case neutrinos gain
generic masses from a type II seesaw. Various neutrino mass patterns
could be distinguished by measuring LFV processes as explored in
\cite{mrs1}. The decay $\mu \rightarrow 3e$ is significantly
enhanced in the case of degenerate or inverted-hierarchical masses
compared with that of the normal hierarchy, whereas the rate of
$\mu\rightarrow e\gamma$ is almost insensitive to these mass
patterns. {\it Case 2}. $h^\nu\neq 0$: Neutrino mass matrix takes
the general form (\ref{seesaw}) depending on $h^\nu$ as well. We
recall that most of $f^\nu$ are constrained by several LFV
processes, but $h^\nu$ does not. The neutrino masses are not
constrained by such processes. Thus there is no relationship among
neutrino mass spectra and LFV rates, in comparison to the first
case.

\section{\label{conclus} Conclusions}

Origin 3-3-1 model with RH neutrinos could not generate consistent
neutrino masses, simultaneously gives large LFV muon decays, which
is the same as in simple extensions of the SM \cite{chli}. By
introducing of the Higgs sextet, the neutrino masses are naturally
small induced via a type II seesaw mechanism. The seesaw scale is
one of the 3-3-1 symmetry breaking scales, it is signified in TeV
order if the neutrino Dirac masses are around electron mass. But, it
reaches $10^8$ GeV whether the Dirac masses rise to muon or tauon
mass. The neutrino mass matrix is given in the correct form revising
that in \cite{vasi}.

LFV decays $\mu\rightarrow 3e$ and $\mu\rightarrow e \gamma$ are
mediated only by the doubly-charged Higgs sextet at high seesaw
scales $(10^8-10^{16})$ GeV, but in TeV scale the singly-charged
scalars also contributing to $\mu\rightarrow e \gamma$. These decay
rates are large and comparable to the current experimental bounds.
In limit $h^\nu=0$, the neutrino mass and LFV happen similarly to
the Higgs triplet model. On other cases of $h^\nu\neq 0$, the
neutrino masses are almost not constrained by such lepton flavor
violating processes. \\

\section*{Acknowledgments}

P.V.D. is grateful to Nishina Fellowship Foundation for financial
support. He would like to thank Prof. Y. Okada and Members of Theory
Group at KEK for support and comments, and Prof. C.S. Lim at Kobe
University for discussions. This work was also supported by National
Council for Natural Sciences of Vietnam.

\end{document}